\documentclass[12pt, letterpaper]{article}
\usepackage{graphicx}
\usepackage{amsmath}

\begin{document}

\author{V. V. Prosentsov\thanks{%
e-mail: vitaly.prosentsov@gmail.com} \\
%EndAName
Laan door deVeste 12, 5708 ZZ, Helmond, The Netherlands}
\title{Wave scattering by objects made of small particles with pulsating permittivity}
\maketitle

\begin{abstract}
In this work we study the wave scattering by small dispersionless particles with pulsating refractive index. The scattered fields and their resonance frequencies are calculated by using scalar approximation and exponentially time-dependent permittivity.

In addition, the scattering by single sphere with the pulsating permittivity is calculated numerically for different regimes of the pulsations.

 Our results suggest that the pulsations of the refractive index of the scatterer significantly affect the scattered field: existing resonances shift, additional resonances emerge, and deeps in the light scattering spectrum appear.
\end{abstract}

\section*{Introduction}
Light scattering is very broad and classical topic of Optics, and it is extensively discussed in the
literature \cite{Born}, \cite{Jackson}. The scattering by small particles is one of the subtopics of the light
scattering, and it has many applications in scatterometry, optics of meta materials, and
contamination detection (see, for example \cite{Hulst}, and references therein).

Till recently, the small particles were studied mostly as ones having constant refractive index, while their shapes could be varied broadly \cite{Hulst}-\cite{Chaumet}. In addition, the scattering by the moving particles was well established
topic long ago, and it became fruitful branch of the Optics \cite{Pecora}.

The next next logical step would be to study the wave propagation in time-varying media, and scattering by particles with refractive index varied in time. Indeed, the number of such investigations is surged recently (see, for example \cite{Hayrapetyan}-\cite{Ptitcyn}, and references therein). In the works \cite{Asadchy}-\cite{Ptitcyn}, the scattering by the sphere with time varying refractive index was studied.  In the paper \cite{NKarl}, a system with rapidly shifting refractive index was used for the frequency conversion if the light confined in the nanoresonator meta-atoms. In the work \cite{Martijn} it was shown that the refractive index of the scatterers can be changed in time (by illuminating them with THz radiation, for example) and as result, the permittivity of the particles can be changed with the predefined frequency.

It would be interesting to study the wave scattering by the system of small particles with the pulsating (pulse-like) refractive index, and to compare it with the case when refractive index of the particles is constant in time.

The local perturbation method (LPM) is well suited for the theoretical investigation and
numerical modeling of the scattering by arbitrary objects made of small particles (particles which characteristic size is smaller than incident wavelength).  The LPM was used to study the light scattering by small particles with constant
refractive index \cite{Markel}-\cite{Draine}, the light scattering by the moving particles \cite{DynScatt}, and by the particles with oscillating permittivity \cite{MyArXivPaper}. Below, we will apply the LPM to study the scattering by objects made of small particles with the pulse-like permittivities varying in time.

In this paper we study analytically and numerically the wave scattering by number of small dispersionless particles with exponentially time-dependent optical contrasts. Each particle has its own optical contrast and specific time at which it changes. By using the LPM, the explicit expression for the field scattered by the cluster of the particles with the time dependent refractive indexes is calculated in scalar approximation. For better understanding of the scattering, the resonance frequency and the resonance width are also calculated for single sphere. In addition, the field scattered by small sphere was numerically calculated for different regimes of pulsations.

The obtained results suggest that the pulsating refractive index significantly affects the scattering process: existing resonances shift, additional resonances emerge, and deeps in the light scattering spectrum appear.

\section{The theoretical formalism}
In this section we study the light scattering by cluster of the small dispersionless particles with time-dependent refractive index in scalar approximation. The scalar approximation allows to show the main features of the scattering process avoiding at the same time the complexity of the vector case.

The goal of the study is to investigate the scattering by the small particles with the pulsating permittivity, i.e. when the optical contrasts of the particles appear and disappear in time.

Since any scattering object can be considered as one made of small particles, our approach, in principle, can be used for study of scattering by objects with arbitrary shapes and sizes.

The equation describing the electric field $E$ propagating in the medium has the form
\begin{equation}
\bigtriangleup E(\mathbf{r},t)-\frac{\partial^2}{c^2\partial t^2}\varepsilon(\mathbf{r},t)%
E(\mathbf{r},t)=\frac{4\pi}{c^2}\frac{\partial}{\partial t}J(\mathbf{r},t), %
 \label{nova1}
\end{equation}
where the permittivity $\varepsilon(\mathbf{r},t)$ of the host medium filled with the small and pulsating scatterers is
\begin{equation}
\varepsilon(\mathbf{r},t)=\varepsilon _{h}+\sum_{n, \: \upsilon=1}^{N, \: \Upsilon} %
\Delta\varepsilon _{n \upsilon}f_{n}(\mathbf{r}-\mathbf{r}_{n})f_{n\upsilon}(t),
\label{nova2}
\end{equation}
and
\begin{equation}
\Delta\varepsilon _{n \upsilon} \equiv \varepsilon_{sc,n\upsilon}-\varepsilon _{h}.
\label{nova2aa}
\end{equation}
Here $\Delta$ is the Laplacian operator, $\mathbf{r}$ is the radius vector of the observer, and $t$ is time at the observer's location, $c$ is the light velocity in vacuum, and $J$ is the field source. We note that $\varepsilon _{h}$ is the relative (in respect to vacuum) permittivity of the host medium, $\Delta\varepsilon _{n \upsilon} $ is the amplitude of the optical contrast of the $n$-th scatterer related to time $t_{n\upsilon}$. The function $f_{n\upsilon}(t)$ describes the time dependency of the optical contrast of the $n$-th particle related to the pulsating time $t_{n\upsilon}$ ($\upsilon=1..\Upsilon$), and the function $f_{n}(\mathbf{r}-\mathbf{r}_{n})$ describes the shape of the $n$-th scatterer (with the characteristic size $L_{n}$) as
\begin{equation}
f_{n}(\mathbf{r}-\mathbf{r}_{n}) = \left\{ \begin{array}{cc} 1\text{,} & \text{inside particle } \\
0\text{,} & \text{outside particle.} \end{array}  \right.  %
\label{nova3}
\end{equation}
Here $\mathbf{r}$ and $\mathbf{r}_{n}$ are the radius vectors of the observer and the $n$-th particle respectively.

Since we consider the scattering by small particles (or scattering by objects made of them), the local perturbation method (LPM) is suitable tool to investigate this kind of problem. By applying the LPM approach, we use the following relation between the field $E(\mathbf{r},t)$ and the field $E(\mathbf{r}_{n},t)$ inside the small particle
\begin{equation}
\varepsilon(\mathbf{r},t)E(\mathbf{r},t) \approx \varepsilon(\mathbf{r},t)E(\mathbf{r}_{n},t)
\label{nova5}
\end{equation}
for the field $E(\mathbf{r},t)$ in the Eq. (\ref{nova1}). After this, we integrate the modified Eq. (\ref{nova1})
over time $t$, and make use of the Fourier transform in the frequency domain. Finally, we obtain the following
wave equation for the field $\widetilde{E}(\mathbf{r}, \omega) $ in the space-frequency domain
\begin{eqnarray}
\left(\bigtriangleup + k^2\right) \widetilde{E}(\mathbf{r}, \omega) + \frac{k^2}{\varepsilon_{h}}
\sum_{n, \: \upsilon=1}^{N, \: \Upsilon}  \Delta\varepsilon _{n \upsilon}
f_n(\mathbf{r}-\mathbf{r}_{n})  \times  \notag \\  %
  \int^{\infty}_{-\infty}  \widetilde{E}(\mathbf{r}_n, \omega')  \widetilde{f}_{n \upsilon}(\omega-\omega')
   d\omega' = -i\omega \frac{4\pi}{c^2}\widetilde{J}(\mathbf{r}, \omega), %
\label{nova6}
\end{eqnarray}
where
\begin{equation}
k\equiv \frac{\omega }{c}\sqrt{\varepsilon _{h}}= \frac{2\pi }{\lambda },  \:  k(\omega)L_{n}\ll 1.
\label{nova6a}
\end{equation}
Here $k(\omega)$ is a wave number in the host medium, and $L_{n}$ is the characteristic size of the $n$-th scatterer (small particle).

Note that the Eq. (\ref{nova6}) is approximate one, and it is correct only when the small
scatterers are considered, so the condition $k(\omega)L_{n}\ll 1$ should be always satisfied.  The Eq. (\ref{nova6}) is integro-differential equation, and it can be solved when the function $\widetilde{f}_{n \upsilon}(\omega)$ describing particle's permittivity variation in frequency domain is known.

The exponential type of the permittivity pulsations is very convenient for our study: it can describe both short and long pulsations, and it also guarantees the convergence of the integrals appearing in the calculations. We will use this type of the temporal dependence in our study below.

We select the function $f_{n \upsilon}(t)$ for the n-th particle in the following form
 \begin{equation}
f_{n \upsilon}(t) = e^{-|t-t_{n\upsilon}|/\tau_{n\upsilon}},
\label{nova6b}
\end{equation}
with its Fourier transform
\begin{equation}
\widetilde{f}_{n \upsilon}(\omega) = \frac{1}{2\pi}\int^{\infty}_{-\infty} f_{n \upsilon}(t)
e^{i\omega t} dt =\frac { \tau_{n\upsilon} e^{i \omega t_{n\upsilon}} }{\pi (1+\omega^2 \tau^2_{n \upsilon})}.
\label{nova6c}
\end{equation}
Here $\tau_{n\upsilon}$ is the characteristic duration of the permittivity pulsation. As expected, for the relatively short pulsations ($\omega \tau_{n \upsilon} \ll 1$), the Fourier transform (\ref{nova6c}) is proportional to $\tau_{n \upsilon}$, and for the relatively long pulsations, the Fourier transform (\ref{nova6c}) tends to be a delta function of $\omega$.
Please note that from here on, for the definiteness, we will use only positive pulsating times, i.e. $t_{n\upsilon}>0$.

When the temporal dependency (\ref{nova6b}) is adopted, the solution of the Eq. (\ref{nova6}) can be found as the sum of the incident $\widetilde{E}_{in}$ and the scattered $\widetilde{E}_{sc}$ fields
\begin{equation}
\widetilde{E}(\mathbf{r},\omega) = \widetilde{E}_{in}(\mathbf{r},\omega) +
\widetilde{E}_{sc}(\mathbf{r},\omega),  %
\label{nova7}
\end{equation}
where the incident field $\widetilde{E}_{in}(\mathbf{r},\omega)$ is
\begin{eqnarray}
\widetilde{E}_{in}(\mathbf{r},\omega) \equiv \frac{4 \pi i \omega}{c^2} %
\int_{-\infty }^{\infty }\frac{\widetilde{J}(\mathbf{q},\omega)%
e^{i\mathbf{q\cdot r}}}{q^{2}-k^{2}}d\mathbf{q}, \label{nova7aa} \\%%
\widetilde{J}(\mathbf{q, \omega})\equiv \frac{1}{(2\pi) ^4}\int_{-\infty }^{\infty%
}J(\mathbf{r},t)e^{i\omega t-i\mathbf{q\cdot r}}d\mathbf{r}dt, %
\label{nova8}
\end{eqnarray}
and $\widetilde{E}_{sc}$ is the field scattered by $N$ particles (each particle has $\Upsilon$ pulsations)
\begin{eqnarray}
\widetilde{E}_{sc}(\mathbf{r},\omega)  \simeq  \sum_{n, \: \upsilon=1}^{N, \: \Upsilon}
\frac {\Delta\varepsilon _{n \upsilon}
\Phi _{n}(\mathbf{r}, \omega)} { \exp \left( t_{n\upsilon}/ \tau_{n\upsilon} \right) }
\widetilde{E}(\mathbf{r}_{n}, \omega-i/\tau_{n\upsilon}).
\label{nova9}
\end{eqnarray}
Here the function $\Phi_{n}$ is
\begin{equation}
\Phi _{n}(\mathbf{r}, \omega)\equiv \frac{\omega^2}{c^2} \int_{-\infty }^{\infty }
\frac{\widetilde{f}_{n}(\mathbf{q})e^{i\mathbf{q\cdot (r-r}_{n})}}{q^{2}-k^{2}}d\mathbf{q},
\label{nova10}
\end{equation}
where $\widetilde{f}_{n}(\mathbf{q})$ is the Fourier transform of the function $f_{n}(\mathbf{r})$ and it has the following form
\begin{equation}
\widetilde{f}_{n}(\mathbf{q})\equiv \frac{1}{8\pi ^{3}}\int_{-\infty }^{\infty %
}f_{n}(\mathbf{r})e^{-i\mathbf{q\cdot r}}d\mathbf{r}.  %
\label{nova10aa}
\end{equation}

The scattered field (\ref{nova9}) can be presented in more explicit form for the observer positioned outside of the cluster of the particles when $\mathbf{r} \neq \mathbf{r}_{n}$. In this case, the scattered field has the simplified form
\begin{eqnarray}
\widetilde{E}_{sc}(\mathbf{r},\omega)  =  \frac{\omega^2}{4\pi c^2}
\sum_{n, \: \upsilon=1}^{N, \: \Upsilon}
 \frac{\Delta\varepsilon _{n \upsilon} V_{n}e^{ikR_{n}}}{R_{n}
 \exp \left( t_{n\upsilon}/\tau_{n\upsilon} \right) }
  \widetilde{E}(\mathbf{r}_{n}, \omega-i/\tau_{n\upsilon}),
\label{nova10d}
\end{eqnarray}
where the distance from the observer to the $n$-th particle is
\begin{equation}
R_{n} \equiv |\mathbf{R}_{n}|, \; \mathbf{R}_{n} \equiv  \mathbf{r} - \mathbf{r}_{n} \neq 0.%
\label{nova10a30}
\end{equation}

Note, that the expressions (\ref{nova7})-(\ref{nova9}) form the complete solution of the scattering problem,
and the expressions for the scattered fields (\ref{nova9}) and (\ref{nova10d}) are the main results of this section. These formulae show that the classical 'suite' of the scattering parameters of the particle (optical contrast $\Delta\varepsilon$, volume $V$, and field $\widetilde{E}(\mathbf{r}_n,\omega)$) is extended in our case: the field inside the particle depends on the complex frequency  $\omega-i/\tau_{n\upsilon}$, and, in addition, the factor $\exp(-t_{n\upsilon}/\tau_{n\upsilon})$ appears due to the pulsations of the permittivity.

The formula (\ref{nova10d}) for the scattered field $\widetilde{E}_{sc}(\mathbf{r},\omega)$ suggests that for very short pulsations (still allowed by LPM) when $\tau_{n\upsilon} \sim 1/\omega$, the field inside the scatterer depends on the complex argument $\omega-i/\tau_{n\upsilon}$, and when the ratio $t_{n\upsilon}/\tau_{n\upsilon} \to \infty$, the scattered field disappears.
In another limiting case of very long pulsations when $\tau_{n\upsilon} \to \infty$, the field inside the scatterer depends on real argument $\omega$ only, and when $t_{n\upsilon}/\tau_{n\upsilon} \to 0$, the scattered field (\ref{nova10d}) coincides with one presented in the work \cite{VP} for the statical case.

The formula (\ref{nova10d}) also suggests that the cumulative effect of the multiple short pulsations may be not negligible even for single scattering particle, when $t_{n\upsilon}/\tau_{n\upsilon} \gg 1$. This effect is considered in greater detail in the section where the resonance frequency is calculated.

For completeness, we present also the system of equations for the fields $\widetilde{E}(\mathbf{r}_{n}, \omega)$
and $\widetilde{E}(\mathbf{r}_{n}, \omega-i/\tau_{n\upsilon})$ inside the particles located at the points $\mathbf{r}_{n}$
\begin{eqnarray}
\widetilde{E}(\mathbf{r}_{j},\omega) = \widetilde{E}_{in}(\mathbf{r}_{j},\omega) +
\sum_{n, \upsilon=1}^{N, \: \Upsilon}%
\alpha_{jn\upsilon}(\omega, \tau_{n\upsilon})
\widetilde{E}(\mathbf{r}_{n},\omega-i/\tau_{n\upsilon}), \; (1 \leq j \leq N)%
\label{nova11}
\end{eqnarray}
where the coefficients $\alpha_{jn\upsilon}$ are
\begin{equation}
\alpha_{jn\upsilon}(\omega, \tau_{n\upsilon}) \equiv  \frac {\omega^2}{c^2}
\frac{ \Delta\varepsilon_{n\upsilon} }{\exp \left( t_{n\upsilon}/\tau_{n\upsilon} \right)}
\int_{-\infty }^{\infty} \frac {\widetilde{f}_{n}(\mathbf{q})}{q^2-k^2}e^{i\mathbf{q} \cdot \mathbf{R}_{jn}}d\mathbf{q},  %
\label{nova12} %
\end{equation}
and
\begin{equation}
\mathbf{R}_{jn} \equiv \mathbf{r}_{j}-\mathbf{r}_{n}.
\label{nova14} %
\end{equation}
Here we used condition that permittivity of each scattering particle is different and it can be modified independently.

As formulae (\ref{nova7}), (\ref{nova9}),  and (\ref{nova10d}) suggest, in order to calculate the scattered
field $\widetilde{E}_{sc}(\mathbf{r},\omega)$, we need to know the fields $\widetilde{E}(\mathbf{r}_{n}, \omega- i/\tau_{n\upsilon})$ inside the $n$-th particle at the complex frequency $\omega-i/\tau_{n\upsilon}$.
To find these fields, it will be required to solve the system of linear difference equations (\ref{nova11}) with respect to the unknown fields $\widetilde{E}(\mathbf{r}_{n}, \omega -i/\tau_{n\upsilon})$  inside the scattering particles.

Finally, it is worth to compare the obtained results with the static case when there is no modification of
the optical contrast, i.e. when $\tau_{n1} \to \infty$. In these cases, for
the scattered fields (\ref{nova10d}) and for the fields inside the particles (\ref{nova11}) we get respectively
\begin{eqnarray}
\widetilde{E}_{sc}(\mathbf{r},\omega)  = \sum_{n=1}^{N} (\varepsilon _{sc,n1}-\varepsilon_{h}) %
\Phi _{n}(\mathbf{r}, \omega) \widetilde{E}(\mathbf{r}_{n}, \omega), \label{nova20a} \\%%
 \widetilde{E}(\mathbf{r}_{j},\omega) = \widetilde{E}_{in}(\mathbf{r}_{j},\omega) + \sum_{n=1}^{N}%
\alpha_{jn1}(\omega) \widetilde{E}(\mathbf{r}_{n},\omega). \label{nova20b} %
\end{eqnarray}
Note that the expressions (\ref{nova20a}) and (\ref{nova20b}) reproduce the relevant formulae for the
scattering by cluster of small particles (see, for example \cite{VP}).

\section{Analysis for single particle with pulsating refractive index}

In this section we apply the formulae obtained previously for the scattering by single small sphere with the pulsating permittivity. Here we calculate explicitly the field scattered by the sphere, its resonance frequency, and the resonance width.

\subsection{Preliminary notes}

Let us take a closer look at the coefficient $\alpha_{jn\upsilon}(\omega, \tau_{n\upsilon})$ for single scattering particle (N=1). We assume that the spherical particle of the radius $L_1$ is located at the point $\mathbf{r}_{1}$, and temporal behaviour of its permittivity is described by the exponential profile (\ref{nova6b}). Note, that only positive pulsating times will be used in our study, i.e. $t_{1\upsilon}>0$.

Below we will consider two limiting cases important for the scattering.

\textbf{I.} Relatively long pulsations, when $t_{1\upsilon}/\tau_{1\upsilon}  \ll 1$. In this case the coefficient $\alpha_{11\upsilon}$ is
\begin{eqnarray}
\alpha_{11\upsilon}(\omega, \tau_{1\upsilon}) \approx  k^2L_1^2
\frac{\Delta\varepsilon_{1\upsilon}} {2 \varepsilon_h}, \: ( t_{1\upsilon}/\tau_{1\upsilon}  \ll 1, \ kL_1 \ll 1).
\label{nova21b}
\end{eqnarray}
 The expression (\ref{nova21b}) suggests that the coefficient $\alpha_{11\upsilon}$ can be, in principle, not small (of order of unity) when $\Delta\varepsilon_{1\upsilon} \geq 10^2$.

 We note here also that the condition $\tau_{1\upsilon} \gg t_{1\upsilon}$ in fact, restricts the number of the flashes to only one, because $t_{12} \geq t_{11}+\tau_{11}+\tau_{12} > \tau_{12}$ that contradicts to the initial condition.

\textbf{II.} Relatively short pulsations, when $t_{1\upsilon}/\tau_{1\upsilon}  \gg 1$. In this case the coefficient $\alpha_{11\upsilon}$ is
\begin{eqnarray}
\alpha_{11\upsilon}(\omega, \tau_{1\upsilon}) \approx  k^2L_1^2
\frac { \Delta\varepsilon_{1\upsilon}e^{-t_{n\upsilon}/\tau_{n\upsilon} }}
{ 2 \varepsilon_h }, \: ( t_{1\upsilon}/\tau_{n\upsilon}  \gg 1, \ kL_1 \ll 1).
\label{nova22b}
\end{eqnarray}
The expression (\ref{nova22b}) suggests that the coefficient $\alpha_{11\upsilon}$ is very small, unless $\Delta\varepsilon_{1\upsilon} \geq 10^4$. However, the coefficient should not be ignored due to its smallness, because the number of the pulsations $\Upsilon$ can be large, and in this case their cumulative effect can be significant, and it should be taken into account.

Note, that for the flash with duration $\tau_{1\upsilon}=10$ fs, for example, the frequency $1/\tau_{1\upsilon} = 10^{14} $ Hz, and this frequency also has to satisfy the LPM condition $|\omega_{max}\sqrt{\varepsilon}/c -i/\tau_{1\upsilon})|L \ll 1$. This means that there is upper limit for the frequency $1/\tau_{n\upsilon}$, which is $1/\tau _{n\upsilon}  \leq \omega_{max}$.

\subsection{The field inside the particle and the scattered field}

In order to find the fields inside the scattering particles we need to solve the system of linear difference equations (\ref{nova11}).  In our case, when $n=1$ and $\upsilon$ is arbitrary, we need to solve only one equation which we represent in the following form
\begin{equation}
\widetilde{E}(\mathbf{r}_{1},\omega) = \widetilde{E}_{in}(\mathbf{r}_{1},\omega) +
\sum_{\upsilon=1}^{\Upsilon} \alpha_{11\upsilon}(\omega, \tau_{1\upsilon})
\widetilde{E}(\mathbf{r}_{1},\omega-i/\tau_{1\upsilon}).
\label{nova22df}
\end{equation}
 It should be noted here, that even in this oversimplified case of single scatterer, the exact solution of the equation (\ref{nova22df}) is difficult to find in a form suitable for analytical analysis. The solution can be found numerically, however, due to divergence near resonance, this way is a challenge also.

From another side, approximate but yet informative expression for the field inside the particle can be found when we use the relation
\begin{equation}
\widetilde{E} (\mathbf{r}_{1}, \omega-i/\tau_{1\upsilon}) \approx \widetilde{E} (\mathbf{r}_{1}, \omega),
\label{nova22hk}
\end{equation}
which in turn is possible when the following condition is satisfied
\begin{equation}
\omega \gg 1/\tau_{1\upsilon}.
\label{nova22kl}
\end{equation}
Taking into account the relation (\ref{nova22hk}) and the condition (\ref{nova22kl}), we can find the field inside the particle in this zero approximation as
\begin{eqnarray}
\widetilde{E} (\mathbf{r}_{1}, \omega)  \approx
\frac {\widetilde{E}_{in}(\mathbf{r}_1, \omega)} { D(\omega, \tau_{1\upsilon}) },  \:  (\omega \gg 1/\tau_{1\upsilon})
\label{nova23}
\end{eqnarray}
where the denominator $D$ has the form
\begin{eqnarray}
D(\omega, \tau_{1\upsilon}) = 1 - \sum_{\upsilon=1}^{ \Upsilon} \alpha_{11\upsilon}(\omega, \tau_{1\upsilon}).
\label{nova23a}
\end{eqnarray}

The approximate expression (\ref{nova23}) already suggests that the resonance of the field inside the pulsating particle is possible when $ Re(D(\omega, \tau_{1\upsilon})) \to 0$, and the resonance frequency will depend on the pulsating durations $\tau_{1\upsilon}$, pulsation times $t_{1\upsilon}$, and on the properties of the incident field $\widetilde{E}_{in}(\mathbf{r}_1, \omega) $. The influence of the field $\widetilde{E}_{in}(\mathbf{r}_1, \omega) $ on the resonance frequency is not negligible when the field is a steep function of $\omega$.

When the field inside the particle is known, the scattered field can be calculated in space-frequency domain via the formulae (\ref{nova10d}) and (\ref{nova23}) as
\begin{eqnarray}
\widetilde{E}_{sc}(\mathbf{r},\omega)  \approx
\frac{\omega^2 V_{1}e^{ikR_{1}} }{4\pi c^2 R_{1}}\widetilde{E}_{in}(\mathbf{r}_1, \omega)
\sum_{\upsilon=1}^{\Upsilon}
 \frac{\Delta\varepsilon _{1 \upsilon} \exp( -t_{1\upsilon}/\tau_{1\upsilon})  } {D(\omega, \tau_{1\upsilon})}.
\label{nova27}
\end{eqnarray}
The expression (\ref{nova27}) for the scattered field shows that for the extreme case of very long pulsations ($t_{1\upsilon}/ \tau_{1\upsilon}  \ll 1$) we obtain already known formula for the static scattering by the small particle. For another extreme case of very short pulsations ($t_{1\upsilon}/ \tau_{1\upsilon}  \gg 1$) the scattering is effectively zero, and no resonance scattering is possible. However, when the ratio $t_{1\upsilon}/ \tau_{1\upsilon}  \sim 1$ and $1/\tau_{1\upsilon} < \omega$, the 'grey' area exists with possibility of the resonance scattering and other interesting phenomena (see section with numerically calculated fields scattered by the sphere).

\subsection{The resonance frequency and the resonance width}
It should be noted that the resonance frequency is extremely sensitive to the permittivity pulsations. The approximate expression (\ref{nova27}) describes the field scattered by the small sphere, and this field has maxima at the resonance frequencies which are solutions of the equation
 \begin{equation}
 \partial |\widetilde{E}_{sc}(\mathbf{r}, \omega )| / \partial \omega  = 0.
 \label{nova30}
\end{equation}
Taking into account that the coefficient $\alpha_{11\upsilon}$ for the small sphere is
 \begin{equation}
\alpha_{11\upsilon}(\omega, \tau_{1\upsilon}) =  \frac {\omega^2L_1^2}{2c^2}
\frac{ \Delta\varepsilon_{1\upsilon} }{\exp \left( t_{1\upsilon}/ \tau_{1\upsilon} \right)}
\left( 1+ i\frac{2}{3}kL_1 \right),
\label{nova31}
\end{equation}
and assuming that the incident field is relatively slow varying function of $ \omega$, such that
\begin{equation}
\frac{\partial|E_{in}(\mathbf{r}_1,\omega)|} {\partial \omega} < \frac {|E_{in}(\mathbf{r}_1,\omega)|}{\omega},
\label{nova31ag}
\end{equation}
the resonance frequency approximately is
\begin{equation}
\omega_{r} \approx  \frac{ \sqrt{2}c } {L_1 \sqrt { \sum_{\upsilon=1}^{ \Upsilon}  \Delta\varepsilon_{1\upsilon} \exp(-t_{1\upsilon}/ \tau_{1\upsilon})  } }.
\label{nova32}
\end{equation}

The resonance frequency (\ref{nova32}) is a new result suggesting that $\omega_{r}$ takes into account the permittivity modulations $\Delta\varepsilon_{1\upsilon}$, and the ratio of the flashing time $t_{1\upsilon}$ and the pulsing duration $ \tau_{1\upsilon}$. The expression (\ref{nova32}) suggests that for large ratios when $t_{1\upsilon}/ \tau_{1\upsilon}  \gg 1$, the resonance frequency increases exponentially, and no resonance is possible in the framework of the LPM. From another side, large number of the pulsations (flashes) may decrease the resonance frequency, and this effect will be significant for the pulsations belonging to 'grey' area when $t_{1\upsilon}/ \tau_{1\upsilon}  \sim 1$.
It should be emphasized that the resonance frequency (\ref{nova32}) does not depend on the properties of the incident field in this approximation when the condition (\ref{nova31ag}) is valid.

We note also that for the single flash with $t_{11}/ \tau_{11} \to 0$, the expression (\ref{nova32}) transforms into well known formula for the resonance frequency of the single sphere with constant permittivity  \cite{VP}.

For the function $f(\omega)=1/ \left| D(\omega) \right|$ the resonance width $\xi $ can be estimated by using the following expression
\begin{equation}
\xi  \approx  \frac { 2 \sqrt{3} \left| \operatorname{Im}D(\omega_{r}) \right|}
 { \left|     \frac  {  \partial \operatorname{Re}D(\omega )  }
           { \partial \omega}   \right|_{\omega=\omega_{r}} },
\label{nova35}
\end{equation}
and will use this expression for our analysis. By using the expression (\ref{nova35}) and the expression for the resonance frequency (\ref{nova32}) adapted for the small sphere, we estimate the resonance width as
\begin{equation}
\xi  \approx  \frac { 4 } { \sqrt{3} } \frac{ c \sqrt{\varepsilon_{h}} }{L_1}
\frac { 1 } {  \sum_{\upsilon=1}^{ \Upsilon}  \Delta\varepsilon_{1\upsilon} \exp(-t_{1\upsilon}/ \tau_{1\upsilon})  }.
\label{nova35a}
\end{equation}

The formula (\ref{nova35a}) suggests that the resonance width decreases with the growth of the optical contrasts $\Delta\varepsilon_{1\upsilon}$ and number of flashes $\Upsilon$. The effect of the ratio
$t_{1\upsilon}/ \tau_{1\upsilon} $ was discussed earlier.

\begin{figure}[t]
  \centering
  %\hfill
    \includegraphics[width=0.95\textwidth]
    {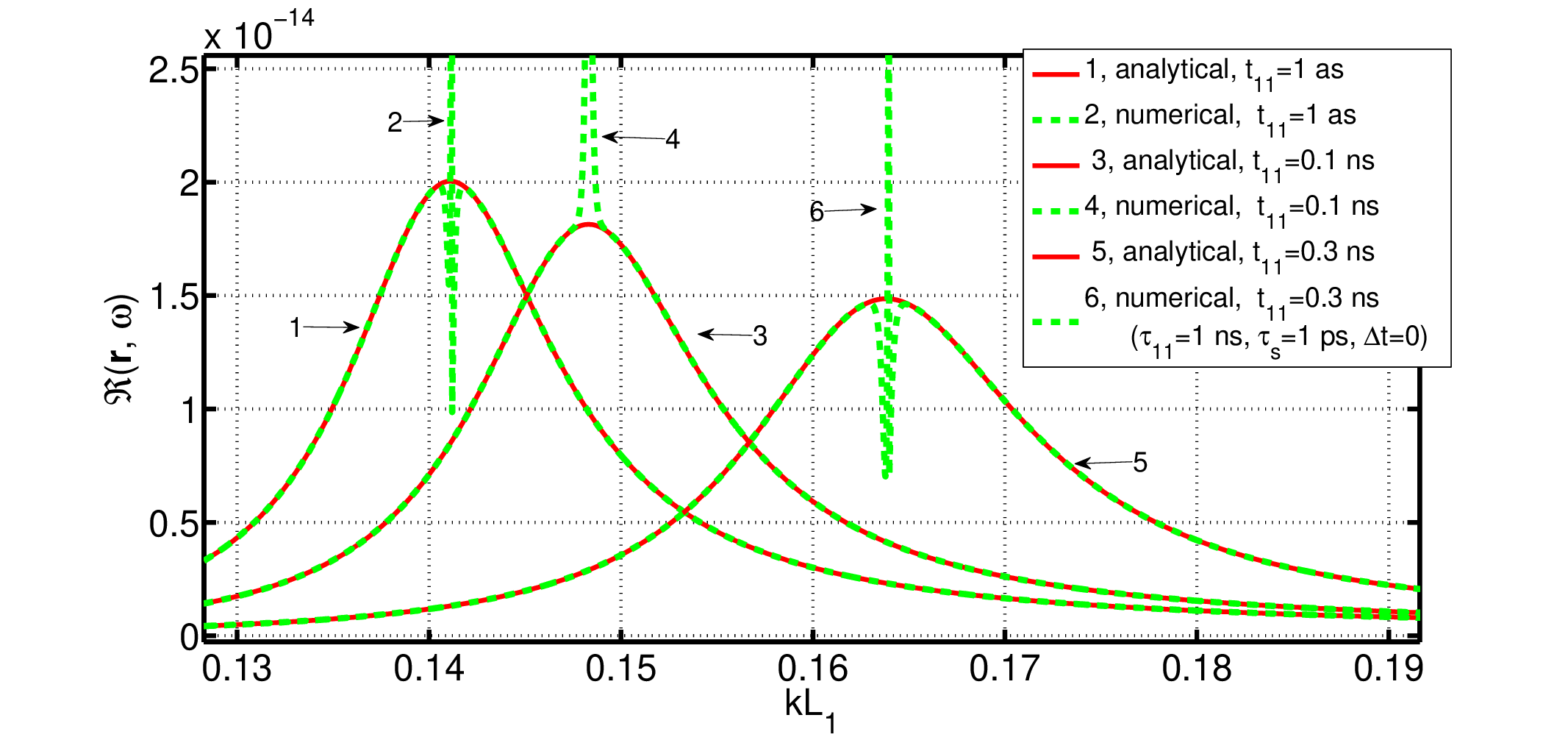}
  \caption{The normalized intensity $\Re(\mathbf{r}, \omega)$ of the field scattered by the small sphere with pulsating permittivity versus normalized frequency $kL_1$.
   The plot suggests that the resonance frequency increases when the ratio $t_{11}/\tau_{11}$ grows. Analytically and numerically calculated intensities are presented for comparison, and they coincide everywhere except resonances. Near the resonances, the numerical results diverge.
  The radius of the sphere is $L_1=20$ nm, its optical contrast is $\Delta\varepsilon_{11}=100$.
  The phase of the incident field was zero by setting $\Delta t =0$, and the source was on for the time $\tau_s=1$ ps. The pulsing duration was set to $\tau_{11}=1$ ns, so the condition $\omega \gg 1/ \tau_{11}$ was very well satisfied. }
  \label{fig1}
\end{figure}

\begin{figure}[t]
  \centering
  %\hfill
    \includegraphics[width=0.95\textwidth]
    {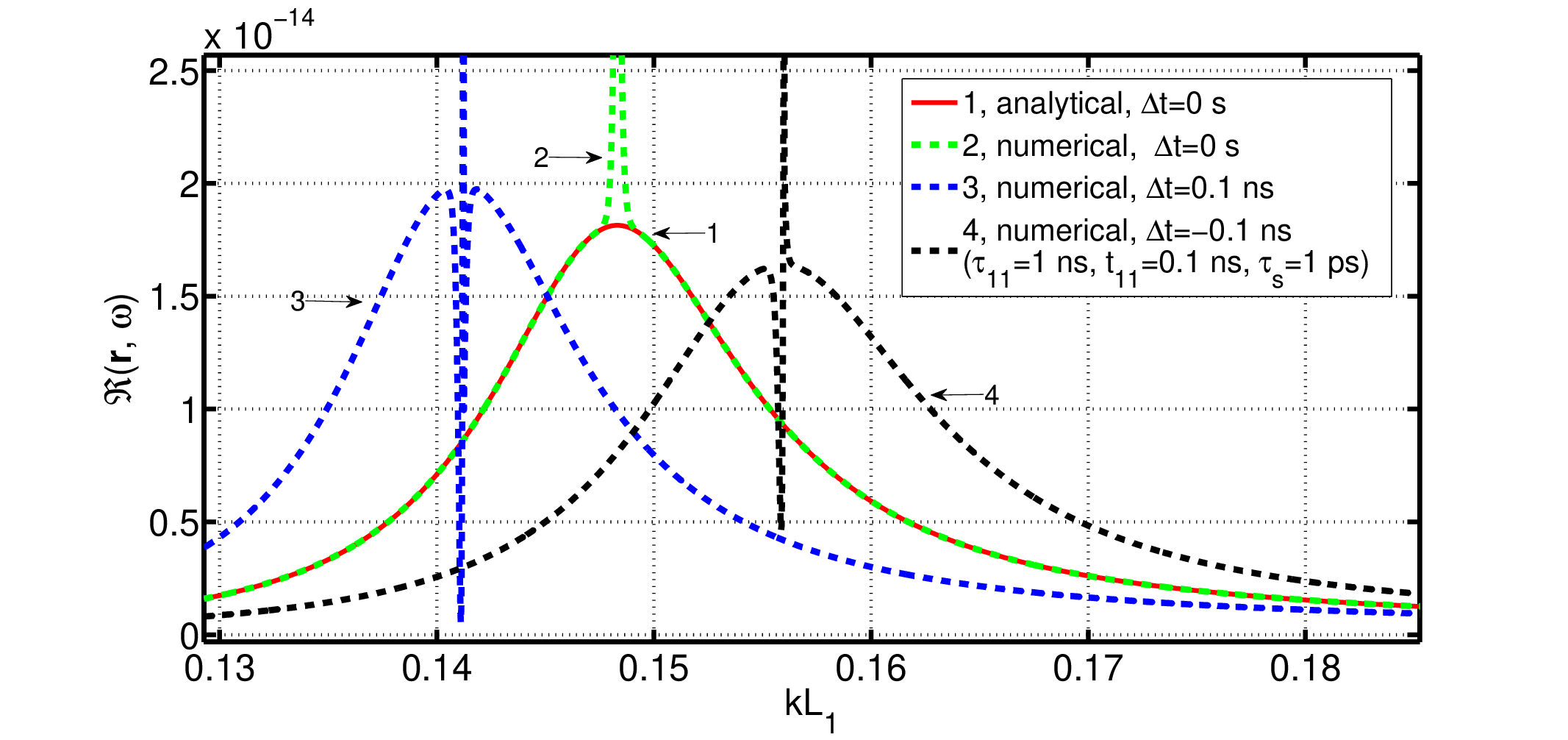}
  \caption{The normalized intensity $\Re(\mathbf{r}, \omega)$ of the field scattered by the small sphere with pulsating permittivity versus normalized frequency $kL_1$.
  The phase of the incident field $\Delta t$ was set to three distinct values: -0.1 ns, 0 ns, and 0.1 ns.  The plot suggests that the resonance frequency decreases or increases when the $\Delta t$ is positive or negative respectively.
  The radius of the sphere is $L_1=20$ nm, and its optical contrast is $\Delta\varepsilon_{11}=100$. The pulsing duration was set to $\tau_{11}=1$ ns, the pulsing time was $t_{11}=0.1$ ns, and the source was on for the time $\tau_s=1$ ps. }
  \label{fig2}
\end{figure}

\begin{figure}[t]
  \centering
  %\hfill
    \includegraphics[width=0.95\textwidth]
    {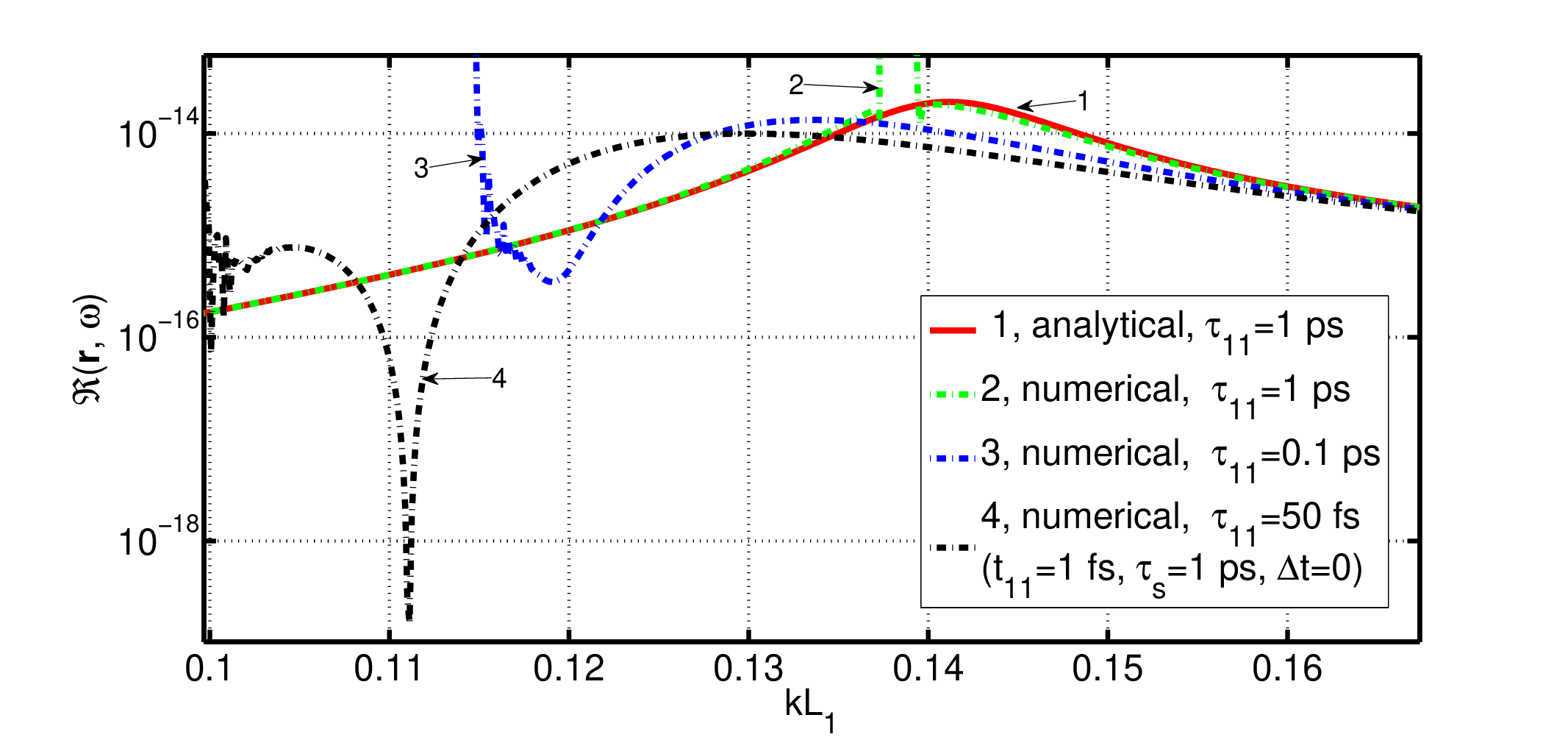}
  \caption{The normalized intensity $\Re(\mathbf{r}, \omega)$ of the field scattered by the small sphere with pulsating permittivity versus normalized frequency $kL_1$. The pulsing durations of the particle were $ \tau_{11}=1 $ ps, 0.1 ps, and 50 fs.
  The plot suggests that at shorter pulsing durations the resonance frequency decreases, and the intensity of the scattered field drops down. For example, at $\tau_{11}=0.1$ ps, and $kL_1 \approx 0.12$ the intensity drop is almost 3 times, while at $\tau_{11}=50$ fs, and $kL_1 \approx 0.11$ the intensity drop is almost 300 times.
  The radius of the sphere is $L_1=20$ nm, and its optical contrast is $ \Delta\varepsilon_{11}=100 $. The pulsing time was $ t_{11}=1$ fs, the source was on for the time $ \tau_s=1 $ ps, and $\Delta t =0$. }
  \label{fig3}
\end{figure}

\section{The numerical modeling of the scattering by single sphere}

In our case, the analytical calculation of the field inside single particle is already a challenge, let alone multiple scattering. Thats is why we will combine analytical and numerical approaches: analytical approach for basic understanding of the wave scattering, and numerical one for revealing interesting features of the scattering.

In this section we will present and discuss the results of the numerical modeling of the field scattered by single sphere with pulsating permittivity.

In all the presented examples the permittivity pulsations are governed by the exponential function (\ref{nova6b}), the radius of the sphere is set to $L_1=20$ nm, and its optical contrast is $\Delta\varepsilon_{11}=100$.

The source current $J(\mathbf{r},t)$ generating the incident field $ E_{in}(\mathbf{r},t) $ was selected in the form of the point source positioned at the point $\mathbf{r}=\mathbf{r}_s$ and emitting light at the moment $t=t_s$ for a time $\tau_s$. In this approach, the source current is represented by product of delta function in space domain and by exponential function in time domain as
\begin{equation}
J(\mathbf{r}, t) =  \delta(\mathbf{r} - \mathbf{r}_s)  \exp(-|t-t_s|/\tau_s),
\label{nova50}
\end{equation}
and the resulting incident field $\widetilde{E}_{in}(\mathbf{r}, \omega)$ has the following form
\begin{equation}
\widetilde{E}_{in}(\mathbf{r},\omega)   = \frac{ i \omega \tau_s}{\pi c^2 R_s} \frac{ \exp( i\omega \Delta t ) }{ (1+\omega^2 \tau_s^2) },
\label{nova50ab}
\end{equation}
where
\begin{equation}
R_s \equiv | \mathbf{r} - \mathbf{r}_s  |, \;  \Delta t \equiv R_s \sqrt{\varepsilon_h}/c + t_s.
\label{nova51}
\end{equation}
Here $R_s $ is the distance from the source to a target (observer or particle), and the time shift
$\Delta t $ shows arriving time of the centre of the incident pulse to a target. When the time shift is zero ($ \Delta t =0 $), the phase of the incident field is also zero, and it will not affect the scattering amplitude via complex frequencies.
In all the used examples the distance $R_s $ from the source to a target and the distance $R_1 $ from the scattering particle to observer were equal ($R_s=R_1=1$ m), and the duration of the incident pulse was set to $\tau_s=1$ ps.

Below we present and discuss the normalized intensity $\Re(\mathbf{r}, \omega)$ of the fields scattered by the sphere. The normalized intensity is defined as ratio of intensity of the field scattered by the scatterer $I_{sc}(\mathbf{r}, \omega)$ to the intensity of the incident field $I_{in}(\mathbf{r}, \omega)$
\begin{equation}
\Re(\mathbf{r},\omega)   = \frac{ I_{sc}(\mathbf{r}, \omega) }{ I_{in}(\mathbf{r}, \omega) } =
 \frac{ |\widetilde{E}_{sc}(\mathbf{r}, \omega)|^2 }{ |\widetilde{E}_{in}(\mathbf{r}, \omega)|^2 } .
\label{nova55}
\end{equation}

The scattered fields were calculated by using the general formula (\ref{nova10d}) and by using approximate formula (\ref{nova27}) whenever possible.

The numerically calculated normalized intensity $\Re$ is presented in the Fig. (\ref{fig1}) for the case when $\omega \gg  1/\tau_{11}$ and the pulsations are relatively long $  t_{11}/\tau_{11} \ll 1$. The pulsation time $t_{11}$ was varied from 1 as to 0.3 ns, while the pulsing duration was fixed at $\tau_{11}=1$ ns, and $\Delta t =0$.
In these particular cases, the approximated formula (\ref{nova23}) for the field inside the particle was used and it was verified by numerical solution of the difference equation (\ref{nova11}).
The behaviour of the intensities in the Fig. (\ref{fig1}) suggest that under the used conditions, the resonance frequency of the field scattered by single sphere increases when the ratio $ t_{11}/\tau_{11}$ grows. This conclusion immediately follows also from the analytical expression for the resonance frequency (\ref{nova32}), however the relation (\ref{nova23}) should have been justified first.

The numerically calculated normalized intensities $\Re$ are presented in the Fig. (\ref{fig2}) for the case of relatively long pulsations and $\omega \gg  1/\tau_{11}$ with addition that $ \Delta t$ of the incident field was varied from $-0.1$ ns to $0.1$ ns. When the phase of the incident field is not zero, the approximate formula (\ref{nova23}) for the field inside the particle is not applicable, and the field inside the scatterer has to be calculated by solving difference equation (\ref{nova11}) numerically.
The plot suggests that in our case, the shift of the resonance frequency depends on the sign of the incident field phase and the phase amplitude: the resonance frequency decreases for the positive phases, and it increases for the negative phases.

Finally, the numerically calculated normalized intensities $\Re$ are presented in the Fig. (\ref{fig3})  for the case of the relatively long pulsations and $1/t_{11} \sim \omega  > 1/\tau_{11}$. The pulsation time was set to $t_{11}=1$ fs, the phase of the incident field was set to zero ($\Delta t =0$), $ \tau_s=1$ ps, and pulsing duration $\tau_{11}$ was varied from $\tau_{11}=1$ ps to 50 fs. At very small times $\tau_{11}$, the approximate formula (\ref{nova23}) is not applicable, and the field inside the scatterer has to be calculated by solving difference equation (\ref{nova11}) numerically.
The plot suggests that the resonance frequency decreases when the pulsation duration $\tau_{11}$ decreases. Also, the results suggest that the intensity of the scattered field may have several minima and maxima. For example, when $\tau_{11}=50$ fs, the intensity has steep and narrow minimum near $kL_1 \simeq 0.111$, while at $kL_1 \simeq 0.105$ the intensity has second maximum.

\section*{Discussions}

The formula (\ref{nova10d}) describes the field scattered by $N$ small particles with time-varying permittivity, and it suggests that the scattering is a complex process depending on the properties of the scatterers (dimensions, optical contrast, time and duration of pulsation), on the properties of the incident field (phase, duration of pulsation, spectrum), and positions of the particles. When the scatterers can be treated as independent ones (due to large distances between the particles, for example), we can use the knowledge obtained in this work for a single scatterer. When the scatterers are not independent, additional study based on the general formula (\ref{nova10d}) is needed.
Due to complexity of the problem, only numerical approach seems most promising for calculation of the fields scattered by cluster of small particles with the pulsating permittivity. This means that some new phenomena are missed in this research.

 We note that the scattering by the cluster of the small particles with time-varying permittivities discussed in this work, is in fact describes the scattering by the complex particles with shape varied in time.

The approximate formula (\ref{nova32}) describes the resonance frequency of the scattered field, showing no dependence on the incident field, because the condition (\ref{nova31ag}) was applied. The numerical results, however, suggest that the resonance frequency of the scattered intensity is sensitive to the phase of the incident field (see Fig. (\ref{fig2})). From the numerical viewpoint, this dependency can be understood as a consequence of the complex frequency used in the calculation and non zero time shift $\Delta t$. This discrepancy can be resolved by taking into account the incident field then the resonance frequency is calculated. In this case, the corrected field (first approximation) inside the particle with single flash ($\upsilon=1$) has the form
\begin{equation}
\widetilde{E} (\mathbf{r}_{1}, \omega)  \approx
 \frac{1 -  2 \alpha(\omega)} { [1 -  \alpha(\omega)]^2 } \widetilde{E}_{in}(\mathbf{r}_1, \omega)  +
 \frac{\alpha(\omega)} { [1 -  \alpha(\omega)]^2 } \widetilde{E}_{in}(\mathbf{r}_1, \omega-i/\tau_{11}),
\label{nova60}
\end{equation}
where
\begin{equation}
\alpha(\omega) \equiv \alpha_{111}(\omega,\tau_{11}),
\label{nova62}
\end{equation}
and the condition $\omega \gg 1/\tau_{11} $ is still valid. By using the field inside the particle (\ref{nova60}) and the incident field (\ref{nova50}), the resonance frequency can be calculated for the ratio $\Re(\mathbf{r},\omega)$ in (\ref{nova55}) as \begin{equation}
\omega'_{r} \approx \omega_{r1} \left( 1 -  \frac {  \Delta t/\tau_{11}  } { 2 + (3\Delta t/\tau_{11})^2/(k_r L_1)^2  } \right),
\label{nova65}
\end{equation}
where
\begin{equation}
\left . \omega_{r1} \equiv  \omega_r \right|_{\upsilon=1}, \  k_r \equiv k|_{\omega=\omega_{r1}}, \  k_r L_1 \ll 1, \   \Delta t/\tau_{11} \ll 1.
\label{nova70}
\end{equation}
The improved formula (\ref{nova65}) correctly predicts the shift of the resonance frequency presented in the Fig. (\ref{fig2}).

Additional feature which is difficult to predict theoretically, is the influence of the pulsing duration of the scatterer on the appearance of additional resonances and deeps in the scattering spectrum (see Fig. (\ref{fig3})). In order to take into account this effect analytically we need to calculate field inside the particle by using high-order approximations (far beyond zeroth and first ones used in formulae (\ref{nova23}) and (\ref{nova60}) respectively), and this issue is not in scope for this work.

\section*{Conclusions}

The fields scattered by the cluster of small dispersionless particles with exponentially time-dependent permittivity have been studied theoretically by using the local perturbation method in scalar approximation. The resonance width, and the resonance frequencies of the field scattered by the particles have been calculated, and it has been shown that they depend on the properties of the permittivity pulsations and the properties of the incident field.

The intensity of the field scattered by the small sphere with the pulsating refractive index has been numerically calculated for various regimes of permittivity pulsations.

 Our results suggest that the permittivity pulsations significantly affect the field scattered by the particle: existing resonances shift, additional resonances emerge, and deeps in the light scattering spectrum appear.

\begin{equation*}
\end{equation*}

\textbf{Acknowledgments}

Many thanks to my mother Lyudmila for her unconditional support.

I would like to express my gratitude to Dr. D. Mazurenko for the helpful discussions.

\end{document}